\documentclass[12pt,preprint]{aastex}

\shorttitle{secondary fan-spine, dome, filament, and flare}
\shortauthors{Hou et al.}

\begin{document}

\title{A secondary fan-spine magnetic structure in active region 11897}

\author{Yijun Hou\altaffilmark{1,2}, Ting Li\altaffilmark{1,2}, Shuhong Yang\altaffilmark{1,2},
        and Jun Zhang\altaffilmark{1,2}}

\altaffiltext{1}{CAS Key Laboratory of Solar Activity, National Astronomical Observatories,
Chinese Academy of Sciences, Beijing 100101, China; yijunhou@nao.cas.cn; zjun@nao.cas.cn}

\altaffiltext{2}{University of Chinese Academy of Sciences, Beijing 100049, China}

\begin{abstract}
Fan-spine is a special topology in solar atmosphere and is closely related to magnetic null point as well as circular-ribbon
flares, which can provide important information for understanding the intrinsic three-dimensional (3D) nature of solar flares.
However, the fine structure within the fan has rarely been investigated. In present paper, we investigate a secondary
fan-spine (SFS) structure within the fan of a larger fan-spine topology. On 2013 November 18, this large fan-spine structure
was traced out due to the partial eruption of a filament, which caused a circular-ribbon flare in active region 11897. The
extrapolated 3D magnetic fields and squashing factor $Q$ maps depict distinctly this fan-spine topology, its surrounding
quasi-separatrix layer (QSL) halo, and a smaller quasi-circular ribbon with high $Q$ located in the center, which implies
the existence of fine structure within the fan. The imaging observations, extrapolated 3D fields, and $Q$ maps on November
17 show that there indeed exists an SFS surrounded by a QSL, which is enveloped by another QSL-halo corresponding to the
overlying larger dome-shaped fan. Moreover, the material flows caused by the null-point reconnection are also detected
along this SFS. After checking the evolution of the underneath magnetic fields, we suggest that the continuous
emergence of magnetic flux within the central parasitic region encompassed by the opposite-polarity fields results in
the formation of the SFS under the large fan.
\end{abstract}


\section{Introduction}
Solar flares are energetic phenomena in the solar atmosphere, converting dramatic free magnetic energy into kinetic
energy of accelerated particles and thermal energy through magnetic reconnection (Shibata 1999; Priest \& Forbes 2002;
Fletcher et al. 2011; Schmieder et al. 2015). To interpret the mechanism of solar flares, many models and theories have
been proposed, among which the ``CSHKP" model is a well-accepted two-dimensional (2D) flare model (Carmichael 1964;
Sturrock 1966; Hirayama 1974; Kopp \& Pneuman 1976). In this unified model, two ribbons observed in H$\alpha$ and
ultraviolet (UV) or extreme ultraviolet (EUV) wavelengths are the most conspicuous characteristics of solar flares.
They are usually located on both sides of the magnetic polarity inversion line (PIL), and their formation is known as
the secondary process when the accelerated particles propagate downward along the newly formed magnetic field lines
and hit the low solar atmosphere (Aschwanden 2002; Shibata \& Magara 2011). As the magnetic reconnection goes on,
new groups of field lines are successively formed with their altitudes moving up and their footpoints (the flare
ribbons) keep separating from each other perpendicularly to the PIL. Therefore, the morphology and dynamics of flare
ribbons provide a significant amount of information about the connectivity of the magnetic field lines and the
reconnection process during solar flares (Gorbachev et al. 1988; Yurchyshyn et al. 2000; Li \& Zhang 2009;
Liu et al. 2010; Savcheva et al. 2015; Qiu et al. 2017; Li et al. 2017b).

Recent high-resolution observations reveal a special kind of flare exhibiting circular ribbon (Masson et al. 2009; Liu
et al. 2011; Wang \& Liu 2012; Sun et al. 2013; Joshi et al. 2015; Hou et al. 2016a; Song \& Tian 2018), which cannot
be explained by the classic 2D flare model. Although many events have been successively interpreted by the 2D model
(Tsuneta 1996; Joshi et al. 2013; Hou et al. 2016b), a three-dimensional (3D) extension of the standard flare model is
necessary for the intrinsic 3D nature of solar flares. For example, the slip-running reconnection model (Aulanier et al.
2006, 2007; Janvier et al. 2013) is constantly developed to interpret the slipping motions of flare loops (Dud{\'{\i}}k
et al. 2014, 2016; Li \& Zhang 2015; Zheng et al. 2016; Jing et al. 2017), and the 3D null-point reconnection process
in fan-spine topology (Lau \& Finn 1990; Antiochos 1998; Priest \& Pontin 2009; Wyper et al. 2018) could explain the
circular ribbon flares. The fan-spine topology is of particular interest in the 3D magnetic reconnection regime, and
usually arises when a new dipole emerges into a pre-existing field (T{\"o}r{\"o}k et al. 2009). One magnetic patch of
this emerging dipole would become a parasitic core encompassed by the opposite-polarity fields, forming a circular PIL.
Then a fan-spine magnetic structure is formed in the corona, in which a dome-shaped fan corresponds to the closed
separatrix surface dividing two different connectivity domains, and the inner and outer spine field lines meet at a null
point. The outer spine can be open or connected to a remote field. In a circular ribbon flare, null-point reconnection
occurs and accelerates particles, which subsequently flow along the separatrix surface and the spine field into the lower
atmosphere. As a result, the circular ribbon is formed at the intersection of the dome-shaped fan and the lower atmosphere
(Hao et al. 2017; Masson et al. 2017; Xu et al. 2017; Lim et al. 2017). When the outer spine field is connected to a remote
field, a remote brightening ribbon would appear (Reid et al. 2012; Li et al. 2017a; Hernandez-Perez et al. 2017). But in
the case that the outer spine field is open, one can see jets along the spine (Pariat et al. 2009; Wang \& Liu 2012;
Jiang et al. 2015; Zhang et al. 2016a; Hong et al. 2017).

Considerable efforts have also been made in numerical simulations and field extrapolations to study the fan-spine
magnetic structure (T{\"o}r{\"o}k et al. 2009; Pariat et al. 2009, 2010; Vemareddy \& Wiegelmann 2014; Yang et al.
2015; Wyper et al. 2018). Masson et al. (2009) reported a confined flare with a circular ribbon, and found that the
circular ribbon outlines the photospheric mapping of the fan field lines and the spine-related ribbons are elongated.
Aided by the potential field extrapolation and magnetohydrodynamics simulation, the authors further suggested that
the fan and spine separatrices are embedded in extended quasi-separatrix layers (QSLs). The extended shape of QSL
surrounding the spine field lines is consistent with the observed elongated spine-related ribbons. The QSL surrounding
the fan separatrix surface is also described as a large QSL-halo with a infinite thickness (Pontin et al. 2016).
In addition, the fan-spine topology in which the outer spine field is open outward was also analyzed to model solar
jets (Pariat et al. 2009; Wyper et al. 2017, 2018; Kumar et al. 2018). To investigate the formation of fan-spine
configurations in the corona, T{\"o}r{\"o}k et al. (2009) modeled the emergence of a twisted flux rope into a
potential field arcade that overlies a weakly twisted coronal flux rope. The authors suggested that a two-step
reconnection process at the null point eventually yields a fan-spine configuration above the emerging dipole.

Despite a large amount of research on fan-spine magnetic topology, fine structures within the dome-shaped fan have
rarely been studied. Kumar et al. (2015) reported the formation and eruption of a small flux rope under a
fan-spine topology by using high-resolution observations from the Goode Solar Telescope (GST) and \emph{Interface
Region Imaging Spectrograph} (\emph{IRIS}; De Pontieu et al. 2014). Hou et al. (2016a) briefly reported the
existence of a secondary fan-spine structure within a dome-shaped magnetic field that belongs to a larger fan-spine
topology. In present paper, by analyzing the high-resolution data from the \emph{IRIS} and \emph{Solar Dynamics
Observatory} (\emph{SDO}; Pesnell et al. 2012), we further study this secondary fan-spine structure in several
aspects, such as magnetic topology, kinematic characteristics, and formation process. The remainder of this paper
is organized as follows. Section 2 describes the observations and data analysis taken in our study. In Section 3,
we investigate the secondary fan-spine structure and present the results of data analysis in detail. Finally, we
summarize the major findings and discuss the results in Section 4.

\section{Observations and Data Analysis}
Based on the observations from \emph{IRIS} and \emph{SDO}, we focus on a fan-spine structure in the complex NOAA
active region (AR) 11897. On 2013 November 18, this fan-spine topology was observed due to the partial eruption of a
filament inside the dome-shaped fan. During the initial formation stage of the fan-spine structure around November 17,
a secondary fan-spine structure was detected within the dome-shaped fan magnetic topology. The Atmospheric Imaging
Assembly (AIA; Lemen et al. 2012) on board \emph{SDO} observes the full solar disk successively in nine (E)UV passbands
with a cadence of (12 s) 24 s and a spatial sampling of 0.{\arcsec}6 pixel$^{-1}$. The \emph{SDO}/Helioseismic and
Magnetic Imager (HMI; Schou et al. 2012) provides one-arcsecond resolution full-disk line of sight (LOS) magnetograms,
Dopplergrams, and intensitygrams every 45 s, and photospheric vector magnetograms at a cadence of 720 s (Hoeksema et al.
2014). To investigate the fan-spine structure on November 18, we use the data of AIA 1600 {\AA}, 304 {\AA}, 131 {\AA},
HMI LOS magnetograms, and the HMI vector data product called Space-weather HMI Active Region Patches (SHARP; Bobra et al.
2014). For the vector magnetic fields, the data of HMI AR patches are deprojected to the heliographic coordinates with
a Lambert (cylindrical equal area) projection method (Gary \& Hagyard 1990; Thompson 2006; Sun et al. 2012). The AIA
131 {\AA}, 1700 {\AA} images, HMI LOS magnetograms, and SHARP data are employed to study the fine structure
within the fan-spine dome-shaped magnetic topology during its initial formation stage from November 16 to 17.

In addition, \emph{IRIS} was pointed at AR 11897 on November 17 and clearly observed the secondary fan-spine structure
within the fan from 12:58:30 UT to 13:47:02 UT. Meanwhile, the spectral data were taken in a very large dense 64-step
raster mode with a step scale of 0.{\arcsec}35 and a step cadence of 5 s. Here we mainly use a series of \emph{IRIS}
slit-jaw images (SJIs) of 1400 {\AA} with a cadence of 10 s, a pixel scale of 0.{\arcsec}166, and a field of view (FOV)
of 167{\arcsec} $\times$ 174{\arcsec} to investigate the secondary fan-spine structure in detail. The 1400 {\AA} channel
contains emission from the Si IV 1394/1403 {\AA} lines formed in the transition region, including the contributions from
UV continuum emission formed in the lower chromosphere as well. The employed data are all level 2, where dark current
subtraction, flat field, geometrical, and orbital variation corrections have been applied (De Pontieu et al. 2014). For
the spectroscopic analysis, we also employ Si IV 1402.77 {\AA} line which is formed in the middle transition region
with a temperature of about 10$^{4.9}$ K. The nearby cold chromospheric S I 1401.51 {\AA} line is used for the absolute
wavelength calibration (Li et al. 2014; Tian et al. 2014, 2018). Both of double-Gaussian fitting and triple-Gaussian
fitting to the Si IV 1402.77 {\AA} line profile are performed to investigate the kinematic characteristics of the
secondary fan-spine structure.

In order to reconstruct the 3D magnetic fields of the fan-spine structures, we utilize the ``weighted optimization"
method to perform nonlinear force-free field (NLFFF) extrapolations (Wiegelmann 2004; Wiegelmann et al. 2012) on
November 18 and November 17, respectively. The vector magnetograms are preprocessed by a procedure developed by
Wiegelmann et al. (2006) to satisfy the force-free condition before being used as boundary condition. Both NLFFF
extrapolations are performed in a box of 896 $\times$ 512 $\times$ 256 uniformly spaced grid points (325 $\times$
186 $\times$ 93 Mm$^{3}$). Furthermore, we calculate the squashing factor $Q$ (Titov et al. 2002) and twist number
$T_{w}$ (Berger \& Prior 2006) of the extrapolated field through the method developed by Liu et al. (2016). The
squashing factor $Q$ provides the most important information about the magnetic connectivity, and the regions with
high $Q$ denote the locations of the QSLs. The twist number $T_{w}$ measures how many turns two field lines wind
about each other and plays a significant role in identifying a magnetic flux rope without ambiguity.

\section{Results}
\subsection{A Fan-spine Configuration in AR 11897 on November 18}
Figure 1 shows a fan-spine configuration in AR 11897 on November 18 (also see movie1.mov). Around 04:16 UT, a
typical fan-spine configuration consisting of a quasi-circular ribbon (CR), a dome-shaped fan and a set of brightening
spine loops (Sun et al. 2013) was detected in AIA 131 {\AA} images (panel (a)). By comparing the 131 {\AA}
image and HMI LOS magnetogram,
we find that the quasi-circular ribbon is located in the negative-polarity fields encircling a positive magnetic core
(panel (b)). Moreover, the 1600 {\AA} image in panel (c) reveals the existence of two small quasi-parallel ribbons
within this quasi-circular ribbon. Focusing on the evolution of the dome-shaped fan region, we expand the 304 {\AA}
images and show them in panels (d1)-(d3). Before the appearance of the brightening fan-spine structure, a filament
was located under the dome-shaped fan (see panel (d1)). Around 04:04 UT, this filament was partly activated and then
erupted upward (see panel (d2)). Then two small quasi-parallel ribbons appeared below the erupting filament (see
panels (c) and (d3)). Superposing the brightness contours of these two ribbons in LOS magnetogram, we notice that
the north ribbon is rooted in the quasi-circular negative magnetic fields while the south ribbon in the central
positive magnetic patch (see the green curves in panel (b)). Subsequently in 131 {\AA} channel, slipping motion of
brightening loops was observed in the south part of the dome-shaped fan. As a result, the whole fan-spine structure
above the erupting filament was traced out accompanied by a quasi-circular ribbon (see panel (a)).

Based on the photospheric vector magnetic fields at 04:00 UT on 2013 November 18, we extrapolated the 3D structure of
AR 11897 by using NLFFF modeling. Moreover, we calculate the twist number $T_{w}$ and the squashing factor $Q$ of the
reconstructed fields. Figures 2(a) and 2(b) show the fan-spine structure completely from the top view and side
view, respectively. The fan-spine magnetic structure is depicted clearly, and consists of a dome-shaped fan and a set
of field lines around the spine (Kumar et al. 2015). The fan divides the coronal space into two connectivity
domains: the inner and the outer domains. The field lines in the inner domain connect the central positive magnetic
patch with the surrounding negative fields (see the white lines in panels (a)-(b)). In the outer domain, the blue lines
connecting the surrounding negative-polarity fields with remote positive fields. According to the photospheric
distribution of $T_{w}$ and the definition of magnetic flux rope mentioned in Liu et al. (2016), we plot the field
lines across the photosphere where the $\mid$$T_{w}$$\mid$ $\geqq$ 1.0. As a result, we obtain a magnetic flux rope
(see the red twisted structure FR in panels (a)-(b)) under the dome-shaped fan, which corresponds to the partially
erupting filament mentioned in Figure 1. The maximum value of the $T_{w}$ of this flux rope is 1.43, which
does not reach the threshold of kink instability at this moment (1.75, T{\"o}r{\"o}k et al. 2004). Panel (c) exhibits
the distribution of the logarithmic $Q$ on the photospheric boundary of the region outlined by the green
dashed square in panel (a). It is shown that there is a quasi-circular ribbon with high $Q$ overlapping the location
of brightening CR in Figure 1. This high-$Q$ ribbon refers to the intersection of a QSL-halo around the fan separatrix
surface with the lower photospheric boundary. The QSL-halo as well as a null point are also obvious in the $Q$ map
(panel (d)) of a vertical plane along the green cut in panel (c). Inside the quasi-circular high-$Q$ ribbon, a QSL
structure around the flux rope (FQ) is clear in the photospheric $Q$ map. What is worth our special attention is that
there is another smaller quasi-circular ribbon with high Q (SQ) located in the center. This ribbon implies the existence
of fine structure within the fan-spine dome-shaped magnetic topology.

\subsection{A Secondary Fan-spine Structure within the Dome-shaped Fan on November 17}
To find the potential fine structure existing within the fan, we check the \emph{IRIS} observations with higher spatial
resolution. A series of \emph{IRIS} 1400 {\AA} images covering the period from 12:58 UT to 13:47 UT on November 17
distinctly reveal a secondary fan-spine structure within the fan-spine dome-shaped magnetic topology described in
Figures 1 and 2. Figures 3(b)-3(c) show this secondary fan-spine structure in 131 {\AA} and 1400 {\AA} wavelengths
(see movie2.mov). Combining the HMI LOS magnetogram in panel (a), we can see that the large dome-shaped fan is rooted
in the surrounding negative-polarity fields (see the quasi-circular ribbon delineated by yellow dashed curve) and
inside this ribbon, a smaller quasi-circular ribbon lies on the inner ring of positive magnetic fields with
central negative fields (see the blue dashed curves). In panel (d), 1400 {\AA} image of the green square in panel (c)
is zoomed in to show the secondary fan-spine configuration. During the evolution of this secondary fan-spine structure,
constant brightening at the top region of the fan and material flows originating from this region in different directions
were observed as well (see the black arrows).

Because the FOV of \emph{IRIS} spectral data during this period covers part of the secondary fan-spine structure, we
perform the spectroscopic analysis to investigate the kinematic characteristics of this small fan-spine structure and
show the results in Figure 4. Figures 4(a1)-4(a3) display the \emph{IRIS} SJIs of 1400 {\AA} at different times when
the spectrograph slits cross different material flows. The black vertical lines mark the locations of slits at different
times, and the plus symbols mark the positions where the slits cross these flows with different directions (see the
arrows). Panels (b1)-(b3) show the Si IV 1402.77 {\AA} line spectra along the three slits marked in panels (a1)-(a3).
The black plus signs in panels (c1)-(c3) delineate the Si IV 1402.77 {\AA} profiles at positions of blue or red plus
symbols shown in panels (b1)-(b3). The asymmetrical property of these profiles indicates that the single-Gaussian fitting
does not apply to them. Thus, we perform double-Gaussian fitting and triple-Gaussian fitting by using the procedures
``dgf\_1lp.pro" and ``tgf\_1lp.pro" in the Solar Software package (Tian et al. 2011) to derive different components
hidden in these asymmetrical profiles. In panel (c1), the observational profile is fitted by a triple-Gaussian fitting
consisting of three single-Gaussian components, which are respectively denoted by the blue, green, and red dashed
profiles. In panels (c2)-(c3), we perform double-Gaussian fitting, and the results are also plotted by black curves
composed of two single-Gaussian components, which are respectively denoted by the blue and red dashed profiles. The
Doppler velocities (V) of each fitting components are separately given. At 12:58:30 UT, the Doppler shift at blue plus
position consists of one slight redshift signal (5.4 km s$^{-1}$) and two strong blueshift signals (--88.4 km s$^{-1}$
and --54.3 km s$^{-1}$), indicating that two kinds of motions may occur simultaneously (Peter 2010; Kumar et al.
2015, 2016): the rotation of the fan-spine field lines and the plasma flows along the fan-spine structures
(the flows denoted by blue arrow in panel (a1)). Similarly in panels (c2) and (c3), a slight shift signal and a strong
shift signal constitute the profiles. The slight shift signals (13.8 km s$^{-1}$ and 27.1 km s$^{-1}$) may represent
the movement the fan-spine structure, and the strong shift signals (64.4 km s$^{-1}$ and 53.7 km s$^{-1}$) correspond
to the material flows along the structure (see the red arrows in panels (a2) and (a3)).

To unambiguously verify the existence of the secondary fan-spine structure under the large dome-shaped fan and study
its magnetic topology, we reconstruct 3D magnetic field above the AR at 13:00 UT on November 17 and show the results
in Figure 5. Similar to Figure 2, here we display the top view of the 3D structures (Figure 5(a)), the distribution
of the logarithmic $Q$ on the bottom boundary (panel (b)), and the $Q$ map in a vertical plane (panel (c)) across the
secondary fan-spine structure. Apparently, one can see the secondary fan-spine (SFS) under a large fan. The yellow
lines in panel (a) delineate the field lines in the inner domain of this SFS connecting the central negative magnetic
patch with the surrounding positive fields. The red lines represent the field lines in the SFS outer domain connecting
the surrounding positive-polarity fields with remote negative fields. In the $Q$ map of panels (b) and (c), the QSL-halo
of the large dome-shaped fan and the QSL around the SFS surface (SQ) are all clearly visible.

\subsection{Formation of the Fan-spine Topologies}
To understand the formation of the secondary fan-spine structure as well as its overlying large fan-spine magnetic
topology, we examine the evolution of their photospheric magnetic fields (see Figure 6 and movie3.mov). We note that a
new dipole (``Dipole 1" in Figure 6(a)) emerged around 04:08 UT on November 16 at the location where the large fan-spine
structure would appear (see the green rectangle). Then the negative patch of ``Dipole 1" kept moving toward the southeast
and was gradually integrated into pre-existing ambient negative fields on the east side. Meanwhile, the positive patch
of ``Dipole 1" continued to grow, being surrounded by the negative fields, and eventually developed into a parasitic
positive-polarity core (see panel (b)). If nothing else, a typical fan-spine magnetic topology would be formed here.
But in present case, continuous flux emergence started within the central positive fields of the already-formed
fan-spine topology around 07:00 UT on November 17 (see the red square in panel (c)). As a result, an isolated negative
patch was formed during the emergence process. On the east of this negative patch (see the blue contour in panel (d)),
positive flux (see the red contour) consistently emerged and moved toward the isolated patch. Meanwhile, brightening
observed in AIA 1700 {\AA} channel repeatedly appeared around the site marked in panel (d), where the fields with
opposite polarities came closer and disappeared (also see the attached movie3.mov). The intermittent brightening
started from $\sim$08:00 UT on November 17 and continued to $\sim$00:00 UT on November 18, when the isolated negative
patch disappeared completely (panel (f)). Combining the AIA observations with extrapolation results, we notice that the
SFS topology appeared around 10:00 UT on November 17, after the appearances of the isolated negative patch and the
brightening in 1700 {\AA} channel, and disappeared before 00:00 UT on November 18.

To illustrate the formation of the secondary fan-spine structure in detail, we propose a cartoon model and show it in
Figure 7. A large dome-shaped fan is rooted in a quasi-circular negative-polarity fields and divides the coronal space
into two connectivity domains: the inner and the outer domain. The blue curves in Figure 7(a) represent the field lines
in the outer domain of the large fan-spine structure, which connect the surrounding negative-polarity fields with remote
positive fields. The white curves mark the field lines in the inner domain connecting the central positive magnetic patch
with the surrounding negative fields. Under this large fan, yellow curves in panel (b) denote the newly-emerging
fields (EF) within the central positive magnetic fields. The star symbols in panel (c) mark the sites where the
newly-emerging fields interact with part of Dipole 1 (D1) (the white curves with small scales in panels (b) and (c)).
Then the newly-formed fields, part of D1 and EF consist of a special fine structure within the large dome-shaped
fan magnetic topology: a secondary fan-spine structure (SFS in panel (d)).

\section{Summary and Discussion}
By analyzing the high-resolution data from the \emph{IRIS} and \emph{SDO}, we investigate a secondary fan-spine structure
within the fan that belongs to a larger fan-spine topology. On 2013 November 18, this large fan-spine topology was detected
accompanied by a C2.8 two-ribbon flare. The flare was caused by the partial eruption of a filament inside the fan.
The extrapolated 3D fields and $Q$ maps depict clearly this fan-spine topology, its surrounding QSL-halo, and a magnetic
flux rope under the fan corresponding to the erupting filament. Specially, there is a smaller quasi-circular ribbon with
high $Q$ located in the center of the QSL-halo, which implies the existence of fine structure within the dome-shaped fan.
Aided by the imaging observations as well as the extrapolated 3D fields, we find that there was indeed a secondary
fan-spine structure under the fan on November 17. The $Q$ maps show that this secondary fan-spine structure is surrounded
by a QSL, which is enveloped by another QSL-halo corresponding to the overlying larger dome-shaped fan. The spectral
observations show that there are material flows along different directions on the surface of this secondary fan-spine
structure. After checking the evolution of the underneath magnetic fields, we notice that a new dipole emerged into a
pre-existing field on November 16. One magnetic patch of this emerging dipole became a parasitic core embedded in the
opposite-polarity fields, which led to the formation of the large fan-spine magnetic structure. Then within the central
parasitic region, new magnetic flux continuously emerged and apparently reconnected with the surrounding fields,
eventually resulting in the formation of the secondary fan-spine structure within the large dome-shaped fan.

In a circular ribbon flare, the eruption of a filament or flux rope under the dome-shaped fan is often observed. As a
result, two small quasi-parallel ribbons caused by the erupting filament would appear within the quasi-circular ribbon
(Joshi et al. 2015; Liu et al. 2015; Zhang et al. 2016b; Hernandez-Perez et al. 2017; Song et al. 2018; Yang \& Zhang 2018).
This phenomenon suggests the
simultaneous presence of at least two magnetic systems during this kind of flare: a magnetic flux rope (whose eruption
causing the two small quasi-parallel ribbons) and a fan-spine topology (whose fan footprints in the lower atmosphere
performing as a quasi-circular ribbon). The destabilization of the underlying flux rope and its interaction with the
fan-spine structure can cause breakout reconnection at the null point, which could be regarded as the triggering
mechanism of the circular ribbon flare. Jiang et al. (2013, 2014) modeled the torus instability of a twisted flux rope
below a dome-shaped fan and the subsequent breakout reconnection at the null. Li et al. (2018) reported a confined
flare with successive formations of two circular ribbons. The authors proposed a compound eruption model consisting
of two episodes of reconnections and suggested that continuous null-point and slipping reconnection took place
between two filaments and the overlying fan-spine fields. In present work, the extrapolation results show that a
magnetic flux rope was located under the dome-shaped fan. And the top of this flux rope had approached the fan at 04:00
UT on November 18, just several minutes before its partial eruption (see Figure 2). Therefore, we suggest that this
flux rope (i.e., the filament in observation) lost its stability due to some mechanisms, such as its increasing twist
caused by the photospheric shear motion or the alleviation of the constraint from its overlying loops. The rising
filament pushed the overlying loops upward, and magnetic reconnection took place. Then two small ribbons were formed in
the lower solar atmosphere (see Figures 1(c) and 1(d3)). When the twisted flux rope reached the null point, null-point
and slipping reconnection occurred between the twisted field of flux rope and the ambient field in the outer domain of
fan-spine topology, leading to the slipping motion of brightening loops in the south part of the dome-shaped fan (see
movie1.mov).

The formation of a fan-spine topology usually arises when a magnetic dipole emerges into regions of essentially unipolar
fields. The fan-spine configuration results from the relaxation of the pre-existing field after reconnecting with the
emerging flux (Antiochos 1998; T{\"o}r{\"o}k et al. 2009; Liu et al. 2011). During this process, one magnetic patch of
this emerging dipole would become a minority polarity isolated in the surrounding opposite-polarity fields, forming a
quasi-circular PIL. Although rarely reported before, it is not difficult to imagine what would happened when another
emerging dipole is embedded into the central parasitic region mentioned above: the formation of a secondary fan-spine
structure within the large dome-shaped fan. Based on the imaging observations and extrapolated fields, we reveal such a
secondary fan-spine structure within the fan that belongs to a larger fan-spine topology on 2013 November 17. In this
case, there was new magnetic flux continuously emerging into the central positive fields of the already-formed
fan-spine topology. As a result, an isolated negative patch was formed, on the east of which, positive flux consistently
emerged and moved toward this isolated patch. Around the site where the fields with opposite polarities came closer and
disappeared, intermittent brightening was observed in AIA 1700 {\AA} channel. It implied that the successive reconnection
occurred between the isolated negative patch and the surrounding fields (Wang \& Shi 1993; Zhang et al. 2001), which
contributed to the formation of the secondary fan-spine structure. To illustrate this process, we propose a cartoon model
shown in Figure 7.

The imaging and spectral observations from the \emph{IRIS} show that there exist material flows originating from a
brightening region along the surface of the secondary fan-spine structure. The results of spectroscopic analysis shown in
Figure 4 reveal that the plasma flows along the fan surface near the quasi-circular ribbon own strong redshift signals,
and the flows above the fan own obvious blueshift signals. These flows could be the outflows caused by the consisting
null-point reconnection near the apex of the dome-shaped fan (Sun et al. 2013; Zeng et al. 2016). The plasma flows with
redshift signals are moving downward along the fan surface, and the flows with blueshift signals are moving upward along
the outer spine above the fan. Their originating region is probably the site of the magnetic null point, where the
reconnection occurs (Kumar et al. 2016). As a result, these flows provide a robust observational evidence for
the existence of a null point configuration consisting of a fan and a spine.

\acknowledgments{
The authors are grateful to the anonymous referee for valuable suggestions.
The data are used courtesy of \emph{SDO} and \emph{IRIS} science teams. \emph{SDO} is a mission of
NASA's Living With a Star Program. \emph{IRIS} is a NASA small explorer mission developed and
operated by LMSAL with mission operations executed at NASA Ames Research center and major
contributions to downlink communications funded by ESA and the Norwegian Space Centre. This work
is supported by the National Natural Science Foundations of China (11533008, 11790304, 11773039,
11673035, 11673034, 11873059, and 11790300), the Youth Innovation Promotion Association of CAS
(2017078 and 2014043), and Key Programs of the Chinese Academy of Sciences (QYZDJ-SSW-SLH050).
}

{}
\clearpage

\begin{figure}
\centering
\includegraphics [width=0.7\textwidth]{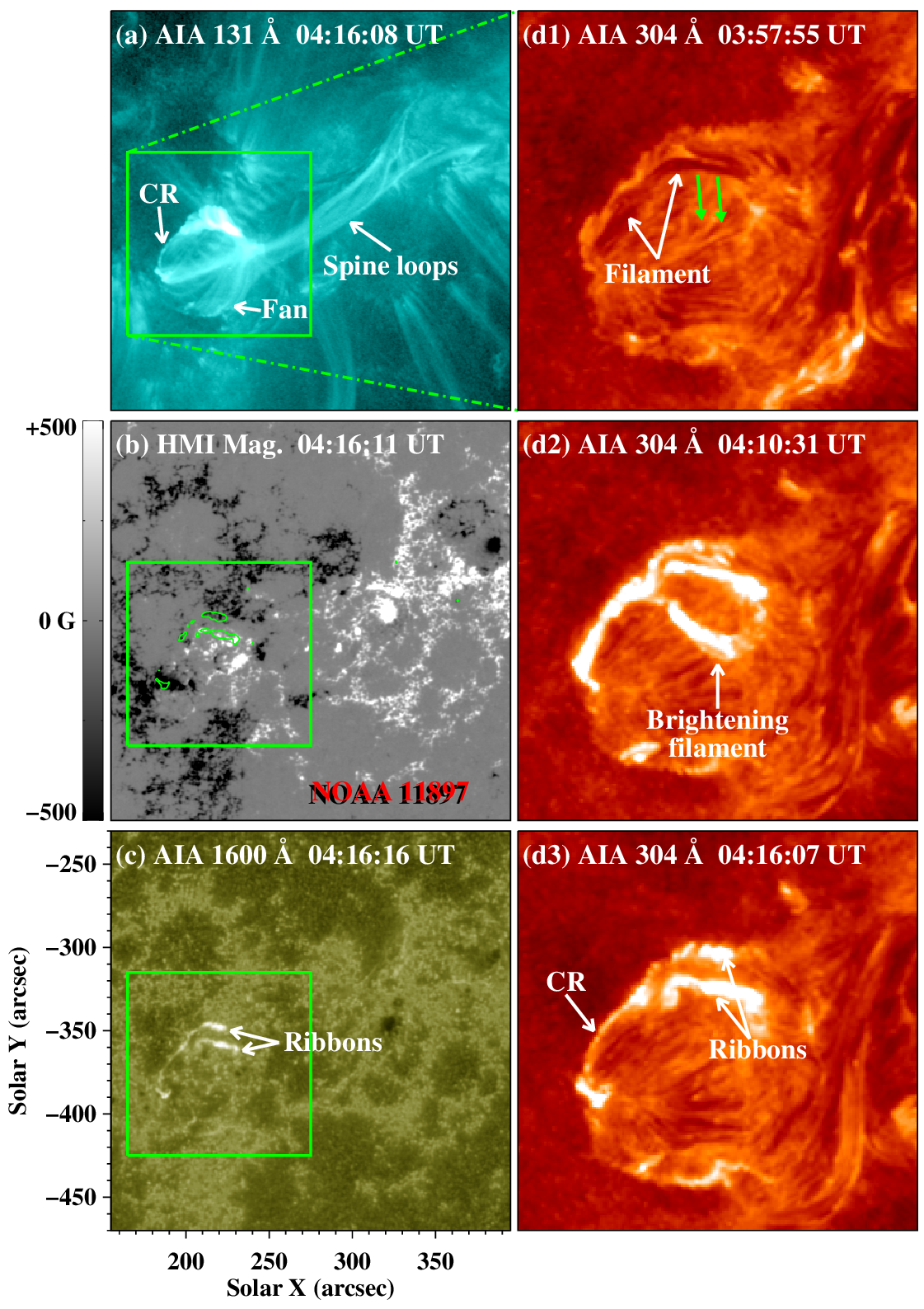}
\caption{Fan-spine structure in AR 11897 on 2013 November 18.
Panels (a)-(c): AIA 131 {\AA} image, HMI LOS magnetogram, and AIA 1600 {\AA} image displaying the overview
of this fan-spine structure and the underlying magnetic fields. The green curves superposed in the magnetogram
of panel (b) are brightness contours of the two quasi-parallel ribbons shown in panel (c).
Panels (d1)-(d3): extended AIA 304 {\AA} images exhibiting a filament under the dome-shaped fan and its
partial eruption process. The field of view (FOV) of these panels is outlined by the green squares in panels
(a)-(c). The green arrows in panel (d1) denote the rising direction of this filament. An animation
(movie1.mov) of 131 {\AA}, 304 {\AA}, and 1600 {\AA} images is available online, which shows the evolution of
the fan-spine structure in AR 11897 from 03:55 UT to 04:44 UT on November 18.
}
\label{fig1}
\end{figure}

\begin{figure}
\centering
\includegraphics [width=0.89\textwidth]{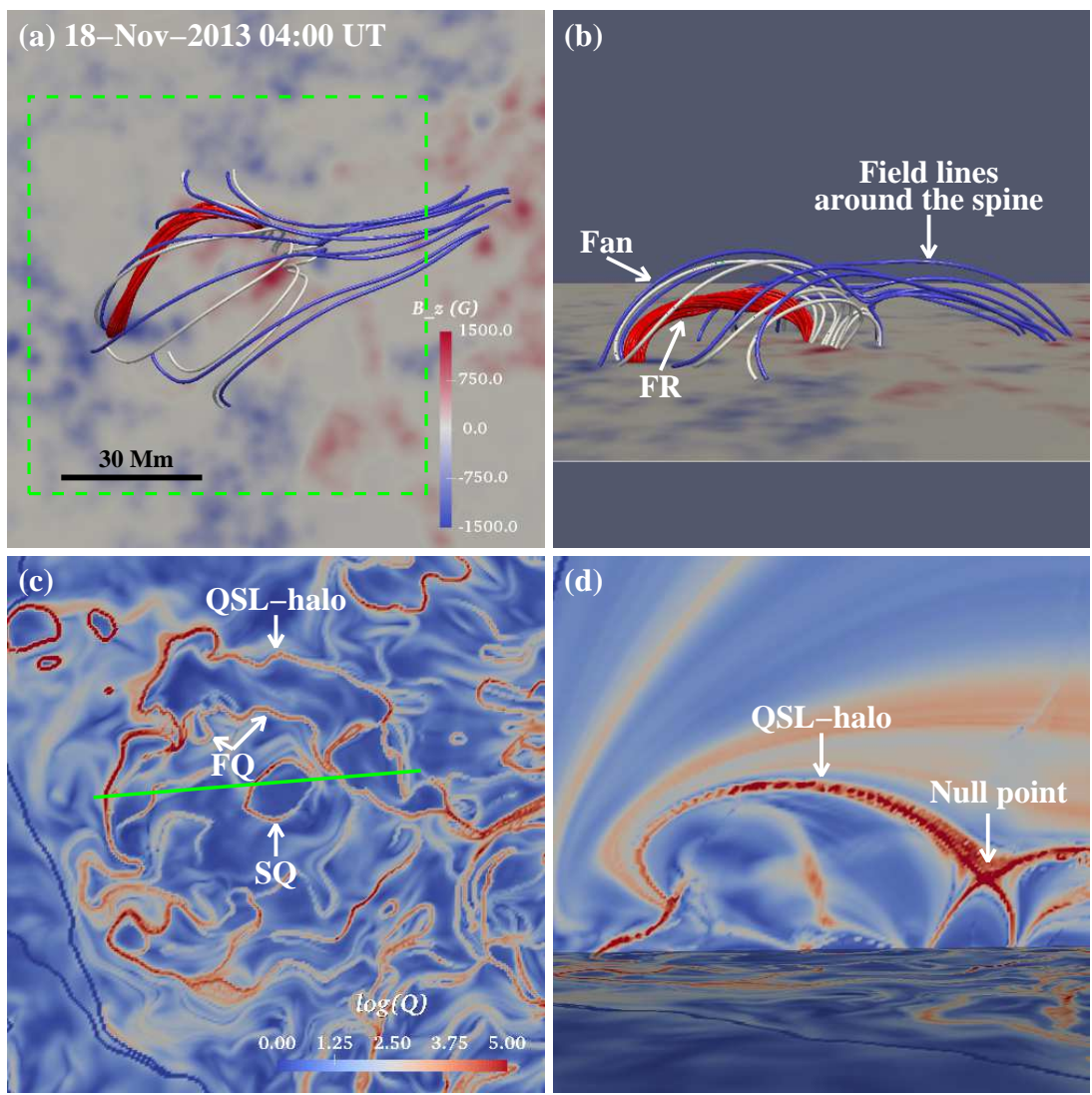}
\caption{Fan-spine configuration revealed by NLFFF modeling at 04:00 UT on November 18.
Panels (a)-(b): top view and side view of the fan-spine structure consisting of a dome-shaped fan and a set of
field lines around the spine. The photospheric vertical magnetic field (B$_z$) is shown as background.
The red twisted field lines represent the magnetic flux rope (FR) located under the dome-shaped fan.
Panel (c): distribution of the logarithmic squashing factor $Q$ on the bottom boundary of the region outlined
by the green dashed square in panel (a). The QSL-halo around the fan and the QSL related to the magnetic flux rope
(FQ) are obvious in this Q map. Specially, a smaller quasi-circular ribbon with high Q is located in the center (SQ).
The FOV of this panel is similar to that of Figures 1(d1)-1(d3).
Panel (d): logarithmic $Q$ map in the vertical plane along the green cut labeled in panel (c).
}
\label{fig2}
\end{figure}

\begin{figure}
\centering
\includegraphics [width=0.89\textwidth]{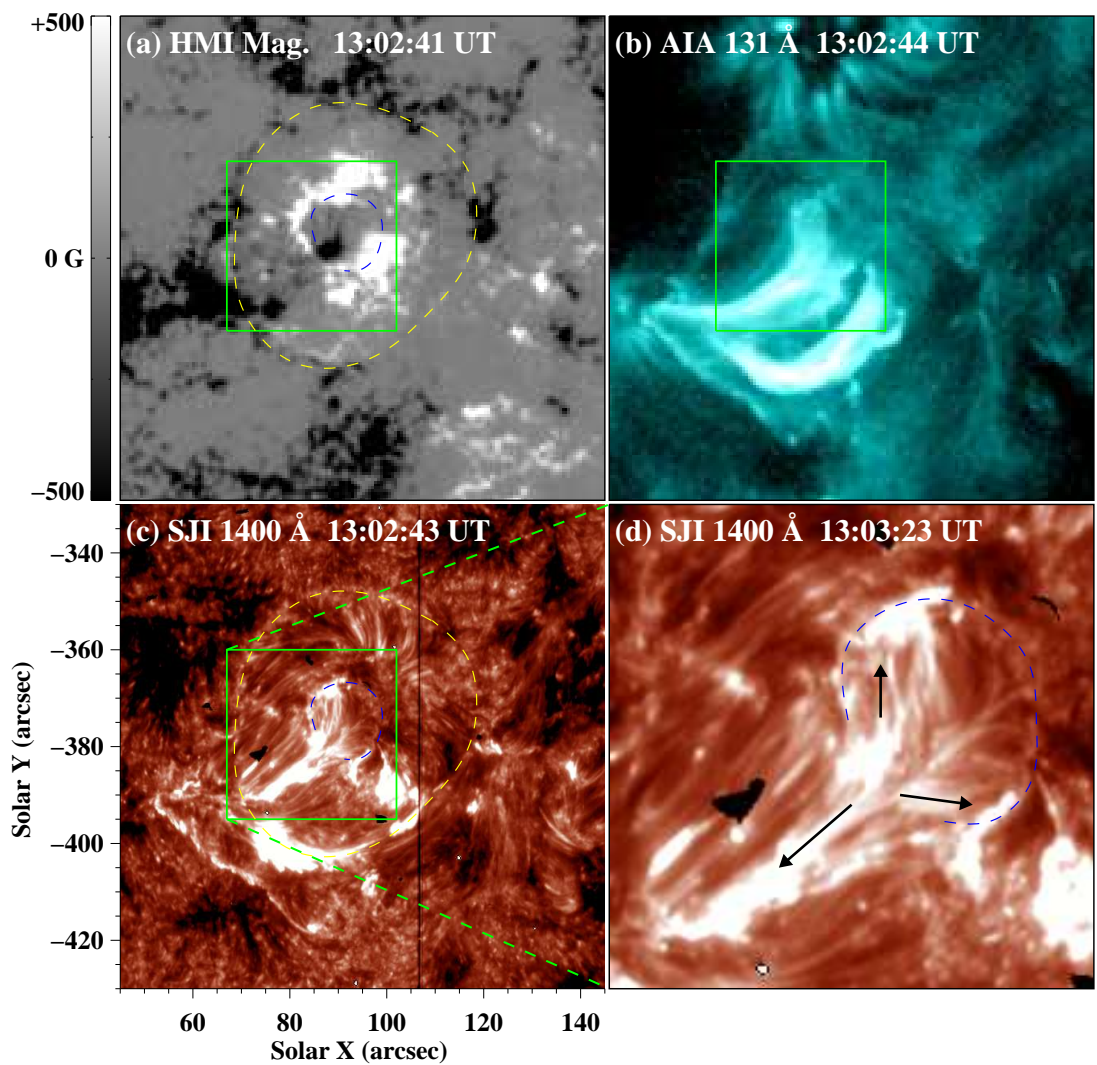}
\caption{Fine structure within the large fan-spine dome-shaped magnetic topology on November 17.
Panels (a)-(c): HMI LOS magnetogram, AIA 131 {\AA} image, and \emph{IRIS} SJI of 1400 {\AA} showing the large
fan-spine structure earlier on November 17. The yellow dashed circles in panels (a) and (c) delineate the
quasi-circular ribbon of this large fan-spine topology. The blue dashed circles denote a smaller quasi-circular
ribbon of the secondary fan-spine structure.
Panel (d): enlarged 1400 {\AA} image of the green square in panel (c) displaying the secondary fan-spine structure.
The black arrows mark different directions of material flows observed in this fine structure.
An animation (movie2.mov) of 1400 {\AA} and 131 {\AA} images is available in the on-line journal, which
displays the secondary fan-spine structure within the large fan-spine dome-shaped magnetic topology from 12:58 UT
to 13:38 UT on November 17.
}
\label{fig3}
\end{figure}

\begin{figure}
\centering
\includegraphics [width=0.92\textwidth]{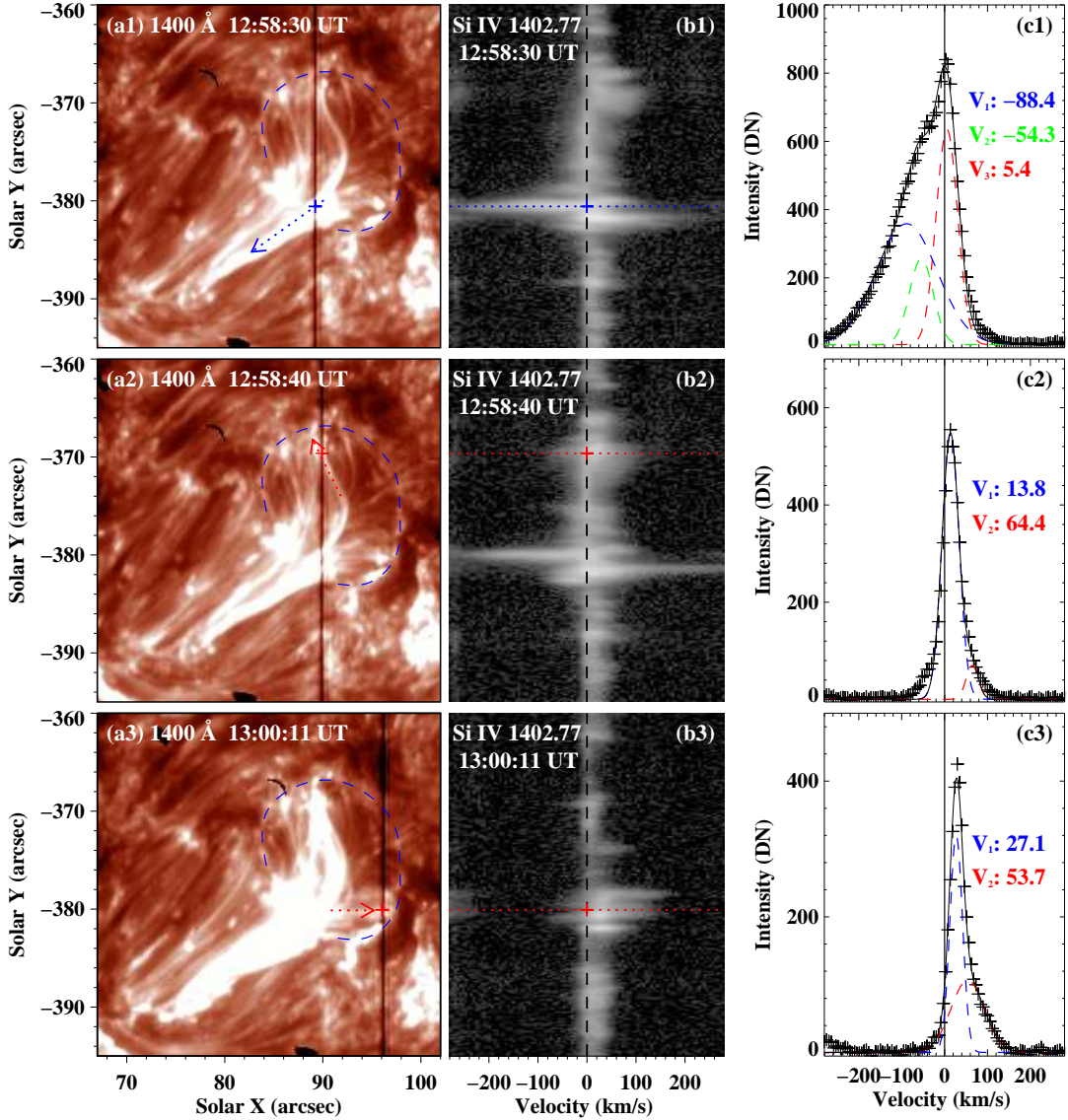}
\caption{Spectral analysis of the kinematic characteristics of the secondary fan-spine structure.
Panels (a1)-(a3): \emph{IRIS} SJIs of 1400 {\AA} displaying different material flows in the secondary fan-spine
structure. The arrows point to the directions of the material flows. The black vertical lines mark the locations
of spectrograph slits at different times.
Panels (b1)-(b3): Si IV 1402.77 {\AA} line spectra along the slits marked in panels (a1)-(a3).
Panels (c1)-(c3): Si IV 1402.77 {\AA} line profiles (black plus signs) and their Gaussian fitting profiles (solid
black curves) at positions of blue or red plus signs shown (dotted lines) in panels (b1)-(b3). The Doppler
velocities (V; km s$^{-1}$) of each single-Gaussian component are respectively given with different colors.
}
\label{fig4}
\end{figure}

\begin{figure}
\centering
\includegraphics [width=0.6\textwidth]{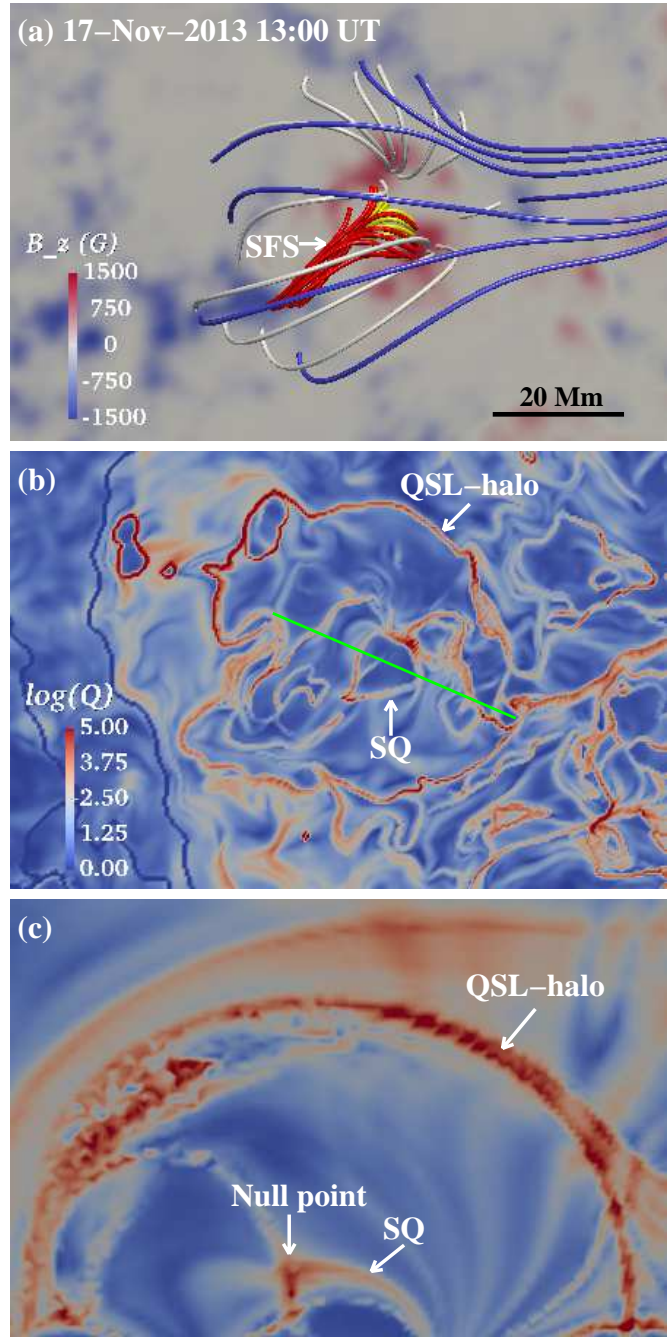}
\caption{Secondary fan-spine (SFS) structure within the dome-shaped fan magnetic topology revealed by NLFFF
extrapolation at 13:00 UT on 2013 November 17. Panels (a) and (b) show these structures from a top view and
the corresponding logarithmic $Q$ map, respectively. The FOV of these panels is similar to that of Figures
3(a)-3(c). Panel (c) displays logarithmic $Q$ distribution in the vertical plane along the green cut denoted
in panel (b), which distinctly depicts the QSL-halo, SQ, and the null point of SFS.
}
\label{fig5}
\end{figure}

\begin{figure}
\centering
\includegraphics [width=0.92\textwidth]{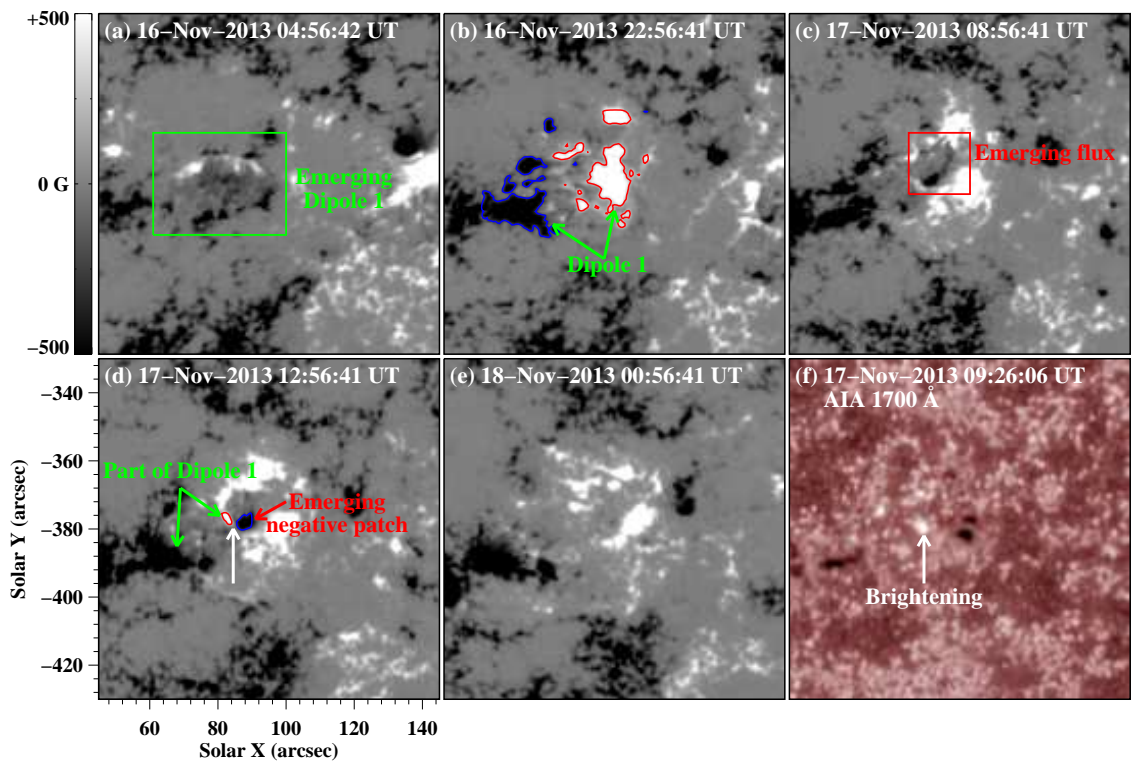}
\caption{Sequence of HMI LOS magnetograms displaying the evolution of magnetic fields in AR 11897. The green
rectangle in panel (a) marks the first emerging dipolar fields (Dipole 1) that resulted in the formation of
the large fan-spine configuration. The red and blue curves in panel (b) are contours of the magnetic fields
of Dipole 1 at +250 and -250 G, respectively. The red rectangle in panel (c) denotes the emerging flux
within the central positive magnetic patch. The white arrow in panel (d) marks the site where the fields
with opposite polarities (see the red and blue contours) came closer and disappeared. The white arrow in panel
(f) denotes the brightening observed in AIA 1700 {\AA} wavelength around the site marked in panel (d).
An animation (movie3.mov) of HMI LOS magnetograms is available online, which shows the evolution of
magnetic fields of AR 11897 from November 16 to November 18. The contours of the brightening shown in 1700 {\AA}
image of panel (f) are superposed on the HMI LOS magnetograms by red curves.
}
\label{fig6}
\end{figure}

\begin{figure}
\centering
\includegraphics [width=0.92\textwidth]{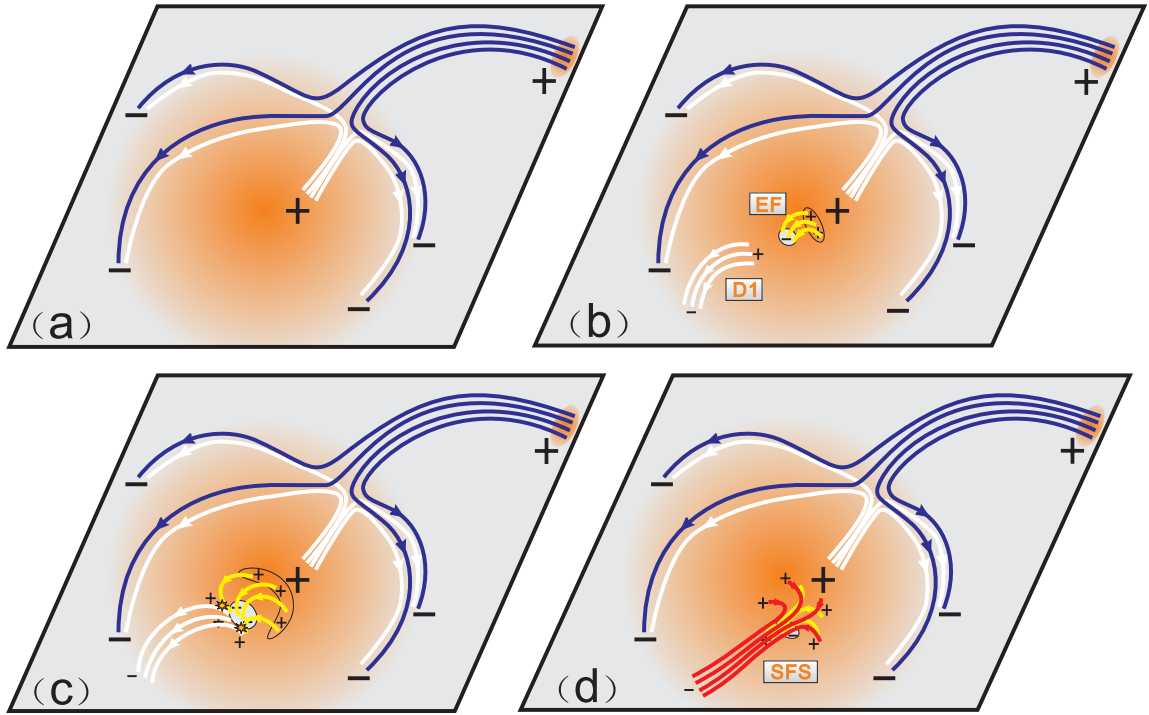}
\caption{
Schematic diagram illustrating the formation of the secondary fan-spine structure within a larger fan-spine
dome-shaped magnetic topology. The orange region in the gray plane represents the positive photospheric magnetic
patch surrounded by negative fields. Newly-emerging fields (EF) and part of Dipole 1 (D1) are marked in
panel (b). Their field lines are denoted by yellow and white curves, respectively. The star symbols in panel (c)
mark the sites where EF interact with part of D1, which form field lines of the secondary fan-spine
(see the red and yellow curves in panel (d)).
}
\label{fig7}
\end{figure}


\begin{thebibliography}{}

\bibitem[Aschwanden(2002)]{2002SSRv..101....1A} Aschwanden, M.~J.\ 2002, \ssr, 101, 1

\bibitem[Antiochos(1998)]{1998ApJ...502L.181A} Antiochos, S.~K.\ 1998, \apjl, 502, L181

\bibitem[Aulanier et al.(2007)]{2007Sci...318.1588A} Aulanier, G., Golub, L., DeLuca, E.~E., et al.\ 2007,
Science, 318, 1588

\bibitem[Aulanier et al.(2006)]{2006SoPh..238..347A} Aulanier, G., Pariat, E., D{\'e}moulin, P., \&
DeVore, C.~R.\ 2006, \solphys, 238, 347

\bibitem[Berger \& Prior(2006)]{2006JPhA...39.8321B} Berger, M.~A., \& Prior, C.\ 2006,
Journal of Physics A Mathematical General, 39, 8321

\bibitem[Bobra et al.(2014)]{2014SoPh..289.3549B} Bobra, M.~G., Sun, X.,
Hoeksema, J.~T., et al.\ 2014, \solphys, 289, 3549

\bibitem[Carmichael(1964)]{1964NASSP..50..451C} Carmichael, H.\ 1964,
NASA Special Publication, 50, 451

\bibitem[De Pontieu et al.(2014)]{2014SoPh..289.2733D} De Pontieu, B.,
Title, A.~M., Lemen, J.~R., et al.\ 2014, \solphys, 289, 2733

\bibitem[Dud{\'{\i}}k et al.(2014)]{2014ApJ...784..144D} Dud{\'{\i}}k, J., Janvier, M.,
Aulanier, G., et al.\ 2014, \apj, 784, 144

\bibitem[Dud{\'{\i}}k et al.(2016)]{2016ApJ...823...41D} Dud{\'{\i}}k, J., Polito, V.,
Janvier, M., et al.\ 2016, \apj, 823, 41

\bibitem[Fletcher et al.(2011)]{2011SSRv..159...19F} Fletcher, L., Dennis, B.~R.,
Hudson, H.~S., et al.\ 2011, \ssr, 159, 19

\bibitem[Gary \& Hagyard(1990)]{1990SoPh..126...21G} Gary, G.~A., \&
Hagyard, M.~J.\ 1990, \solphys, 126, 21

\bibitem[Gorbachev et al.(1988)]{1988SvA....32..308G} Gorbachev, V.~S., Kelner, S.~R., Somov, B.~V., \&
Shvarts, A.~S.\ 1988, \sovast, 32, 308

\bibitem[Hao et al.(2017)]{2017NatCo...8.2202H} Hao, Q., Yang, K., Cheng, X., et al.\ 2017, Nature Communications, 8, 2202

\bibitem[Hernandez-Perez et al.(2017)]{2017ApJ...847..124H} Hernandez-Perez, A., Thalmann, J.~K.,
Veronig, A.~M., et al.\ 2017, \apj, 847, 124

\bibitem[Hirayama(1974)]{1974SoPh...34..323H} Hirayama, T.\ 1974,
\solphys, 34, 323

\bibitem[Hoeksema et al.(2014)]{2014SoPh..289.3483H} Hoeksema, J.~T.,
Liu, Y., Hayashi, K., et al.\ 2014, \solphys, 289, 3483

\bibitem[Hong et al.(2017)]{2017ApJ...835...35H} Hong, J., Jiang, Y., Yang, J., Li, H., \& Xu, Z.\ 2017, \apj, 835, 35

\bibitem[Hou et al.(2016)]{2016A&A...592A.138H} Hou, Y.~J., Li, T., \& Zhang, J.\ 2016a, \aap, 592, A138

\bibitem[Hou et al.(2016)]{2016ApJ...829L..29H} Hou, Y., Zhang, J., Li, T., et al.\ 2016b, \apjl, 829, L29

\bibitem[Janvier et al.(2013)]{2013A&A...555A..77J} Janvier, M., Aulanier, G., Pariat, E., \& D{\'e}moulin, P.\
2013, \aap, 555, A77

\bibitem[Jiang et al.(2013)]{2013ApJ...771L..30J} Jiang, C., Feng, X., Wu, S.~T., \& Hu, Q.\ 2013, \apjl, 771, L30

\bibitem[Jiang et al.(2014)]{2014ApJ...780...55J} Jiang, C., Wu, S.~T., Feng, X., \& Hu, Q.\ 2014, \apj, 780, 55

\bibitem[Jiang et al.(2015)]{2015PASJ...67...78J} Jiang, F., Zhang, J., \& Yang, S.\ 2015, \pasj, 67, 78

\bibitem[Jing et al.(2017)]{2017ApJ...842L..18J} Jing, J., Liu, R., Cheung, M.~C.~M., et al.\ 2017,
\apjl, 842, L18

\bibitem[Joshi et al.(2013)]{2013ApJ...771...65J} Joshi, N.~C., Srivastava, A.~K., Filippov, B., et al.\
2013, \apj, 771, 65

\bibitem[Joshi et al.(2015)]{2015ApJ...812...50J} Joshi, N.~C., Liu, C., Sun, X., et al.\ 2015, \apj, 812, 50

\bibitem[Kopp \& Pneuman(1976)]{1976SoPh...50...85K} Kopp, R.~A.,
\& Pneuman, G.~W.\ 1976, \solphys, 50, 85

\bibitem[Kumar et al.(2016)]{2016ApJ...828...28K} Kumar, P., Innes, D.~E., \&
Cho, K.-S.\ 2016, \apj, 828, 28

\bibitem[Kumar et al.(2018)]{2018ApJ...854..155K} Kumar, P., Karpen, J.~T.,
Antiochos, S.~K., et al.\ 2018, \apj, 854, 155

\bibitem[Kumar et al.(2015)]{2015ApJ...809...83K} Kumar, P., Yurchyshyn, V., Wang, H., \&
Cho, K.-S.\ 2015, \apj, 809, 83

\bibitem[Lau \& Finn(1990)]{1990ApJ...350..672L} Lau, Y.-T., \& Finn, J.~M.\ 1990, \apj, 350, 672

\bibitem[Lemen et al.(2012)]{2012SoPh..275...17L} Lemen, J.~R., Title, A.~M.,
Akin, D.~J., et al.\ 2012, \solphys, 275, 17

\bibitem[Li et al.(2017)]{2017ApJ...836..235L} Li, H., Jiang, Y., Yang, J., et al.\ 2017a, \apj, 836, 235

\bibitem[Li et al.(2014)]{2014A&A...570A..93L} Li, L.~P., Peter, H., Chen, F., \&
Zhang, J.\ 2014, \aap, 570, A93

\bibitem[Li \& Zhang(2009)]{2009ApJ...690..347L} Li, L., \& Zhang, J.\ 2009, \apj, 690, 347

\bibitem[Li et al.(2018)]{2018ApJ...859..122L} Li, T., Yang, S., Zhang, Q., Hou, Y., \& Zhang, J.\ 2018,
\apj, 859, 122

\bibitem[Li \& Zhang(2015)]{2015ApJ...804L...8L} Li, T., \& Zhang, J.\ 2015, \apjl, 804, L8

\bibitem[Li et al.(2017)]{2017ApJ...848...32L} Li, T., Zhang, J., \& Hou, Y.\ 2017b, \apj, 848, 32

\bibitem[Lim et al.(2017)]{2017ApJ...850..167L} Lim, E.-K., Yurchyshyn, V., Kumar, P., et al.\ 2017,
\apj, 850, 167

\bibitem[Liu et al.(2015)]{2015ApJ...812L..19L} Liu, C., Deng, N., Liu, R., et al.\ 2015, \apjl, 812, L19

\bibitem[Liu et al.(2010)]{2010ApJ...721L.193L} Liu, C., Lee, J., Jing, J., et al.\ 2010, \apjl, 721, L193

\bibitem[Liu et al.(2016)]{2016ApJ...818..148L} Liu, R., Kliem, B.,
Titov, V.~S., et al.\ 2016, \apj, 818, 148

\bibitem[Liu et al.(2011)]{2011ApJ...728..103L} Liu, W., Berger, T.~E., Title, A.~M., Tarbell, T.~D.,
\& Low, B.~C.\ 2011, \apj, 728, 103

\bibitem[Masson et al.(2009)]{2009ApJ...700..559M} Masson, S., Pariat, E., Aulanier, G., \&
Schrijver, C.~J.\ 2009, \apj, 700, 559

\bibitem[Masson et al.(2017)]{2017A&A...604A..76M} Masson, S., Pariat, {\'E}., Valori, G., et al.\ 2017,
\aap, 604, A76

\bibitem[Pariat et al.(2009)]{2009ApJ...691...61P} Pariat, E., Antiochos, S.~K., \& DeVore, C.~R.\
2009, \apj, 691, 61

\bibitem[Pariat et al.(2010)]{2010ApJ...714.1762P} Pariat, E., Antiochos, S.~K., \& DeVore, C.~R.\
2010, \apj, 714, 1762

\bibitem[Pesnell et al.(2012)]{2012SoPh..275....3P} Pesnell, W.~D.,
Thompson, B.~J., \& Chamberlin, P.~C.\ 2012, \solphys, 275, 3

\bibitem[Peter(2010)]{2010A&A...521A..51P} Peter, H.\ 2010, \aap, 521, A51

\bibitem[Pontin et al.(2016)]{2016SoPh..291.1739P} Pontin, D., Galsgaard, K., \& D{\'e}moulin, P.\ 2016,
\solphys, 291, 1739

\bibitem[Priest \& Forbes(2002)]{2002A&ARv..10..313P} Priest, E.~R., \& Forbes, T.~G.\ 2002, \aapr, 10, 313

\bibitem[Priest \& Pontin(2009)]{2009PhPl...16l2101P} Priest, E.~R., \& Pontin, D.~I.\ 2009,
Physics of Plasmas, 16, 122101

\bibitem[Qiu et al.(2017)]{2017ApJ...838...17Q} Qiu, J., Longcope, D.~W., Cassak, P.~A., \& Priest, E.~R.\
2017, \apj, 838, 17

\bibitem[Reid et al.(2012)]{2012A&A...547A..52R} Reid, H.~A.~S., Vilmer, N., Aulanier, G., \& Pariat, E.\ 2012,
\aap, 547, A52

\bibitem[Savcheva et al.(2015)]{2015ApJ...810...96S} Savcheva, A., Pariat, E., McKillop, S., et al.\ 2015,
\apj, 810, 96

\bibitem[Schmieder et al.(2015)]{2015SoPh..290.3457S} Schmieder, B., Aulanier, G., \& Vr{\v s}nak, B.\ 2015,
\solphys, 290, 3457

\bibitem[Schou et al.(2012)]{2012SoPh..275..229S} Schou, J., Scherrer,
P.~H., Bush, R.~I., et al.\ 2012, \solphys, 275, 229

\bibitem[Shibata(1999)]{1999Ap&SS.264..129S} Shibata, K.\ 1999, \apss, 264, 129

\bibitem[Shibata \& Magara(2011)]{2011LRSP....8....6S} Shibata, K., \& Magara, T.\ 2011,
Living Reviews in Solar Physics, 8, 6

\bibitem[Song et al.(2018)]{2018ApJ...854...64S} Song, Y.~L., Guo, Y., Tian, H., et al.\ 2018, \apj, 854, 64

\bibitem[Song \& Tian(2018)]{2018ApJ...867..159S} Song, Y., \& Tian, H.\ 2018, \apj, 867, 159

\bibitem[Sturrock(1966)]{1966Natur.211..695S} Sturrock, P.~A.\ 1966, \nat, 211, 695

\bibitem[Sun et al.(2012)]{2012ApJ...748...77S} Sun, X., Hoeksema, J.~T.,
Liu, Y., et al.\ 2012, \apj, 748, 77

\bibitem[Sun et al.(2013)]{2013ApJ...778..139S} Sun, X., Hoeksema, J.~T., Liu, Y., et al.\ 2013, \apj, 778, 139

\bibitem[Thompson(2006)]{2006A&A...449..791T} Thompson, W.~T.\ 2006, \aap, 449, 791

\bibitem[Tian et al.(2014)]{2014ApJ...786..137T} Tian, H., DeLuca, E.,
Reeves, K.~K., et al.\ 2014, \apj, 786, 137

\bibitem[Tian et al.(2011)]{2011ApJ...738...18T} Tian, H., McIntosh, S.~W., De Pontieu, B., et al.\ 2011,
\apj, 738, 18

\bibitem[Tian et al.(2018)]{2018ApJ...854...92T} Tian, H., Yurchyshyn, V., Peter, H., et al.\ 2018,
\apj, 854, 92

\bibitem[Titov et al.(2002)]{2002JGRA..107.1164T} Titov, V.~S., Hornig, G., \& D{\'e}moulin, P.\ 2002,
Journal of Geophysical Research (Space Physics), 107, 1164

\bibitem[T{\"o}r{\"o}k et al.(2009)]{2009ApJ...704..485T} T{\"o}r{\"o}k, T., Aulanier, G., Schmieder, B.,
Reeves, K.~K., \& Golub, L.\ 2009, \apj, 704, 485

\bibitem[T{\"o}r{\"o}k et al.(2004)]{2004A&A...413L..27T} T{\"o}r{\"o}k, T., Kliem, B.,
\& Titov, V.~S.\ 2004, \aap, 413, L27

\bibitem[Tsuneta(1996)]{1996ApJ...456..840T} Tsuneta, S.\ 1996, \apj, 456, 840

\bibitem[Vemareddy \& Wiegelmann(2014)]{2014ApJ...792...40V} Vemareddy, P., \& Wiegelmann, T.\ 2014, \apj, 792, 40

\bibitem[Wang \& Liu(2012)]{2012ApJ...760..101W} Wang, H., \& Liu, C.\ 2012, \apj, 760, 101

\bibitem[Wang \& Shi(1993)]{1993SoPh..143..119W} Wang, J., \& Shi, Z.\ 1993, \solphys, 143, 119

\bibitem[Wiegelmann(2004)]{2004SoPh..219...87W} Wiegelmann, T.\ 2004, \solphys, 219, 87

\bibitem[Wiegelmann et al.(2006)]{2006SoPh..233..215W} Wiegelmann, T., Inhester, B.,
\& Sakurai, T.\ 2006, \solphys, 233, 215

\bibitem[Wiegelmann et al.(2012)]{2012SoPh..281...37W} Wiegelmann, T., Thalmann, J.~K.,
Inhester, B., et al.\ 2012, \solphys, 281, 37

\bibitem[Wyper et al.(2017)]{2017Natur.544..452W} Wyper, P.~F., Antiochos, S.~K., \& DeVore, C.~R.\ 2017, \nat, 544, 452

\bibitem[Wyper et al.(2018)]{2018ApJ...852...98W} Wyper, P.~F., DeVore, C.~R., \& Antiochos, S.~K.\ 2018, \apj, 852, 98

\bibitem[Xu et al.(2017)]{2017ApJ...851...30X} Xu, Z., Yang, K., Guo, Y., et al.\ 2017, \apj, 851, 30

\bibitem[Yang et al.(2015)]{2015ApJ...806..171Y} Yang, K., Guo, Y., \& Ding, M.~D.\ 2015, \apj, 806, 171

\bibitem[Yang \& Zhang(2018)]{2018ApJ...860L..25Y} Yang, S., \& Zhang, J.\ 2018, \apjl, 860, L25

\bibitem[Yurchyshyn et al.(2000)]{2000ApJ...540.1143Y} Yurchyshyn, V.~B., Wang, H., Qiu, J., Goode, P.~R.,
\& Abramenko, V.~I.\ 2000, \apj, 540, 1143

\bibitem[Zeng et al.(2016)]{2016ApJ...819L...3Z} Zeng, Z., Chen, B., Ji, H., Goode, P.~R., \& Cao, W.\ 2016, \apjl,
819, L3

\bibitem[Zhang et al.(2001)]{2001ApJ...548L..99Z} Zhang, J., Wang, J., Deng, Y., \& Wu, D.\ 2001, \apjl, 548, L99

\bibitem[Zhang et al.(2016)]{2016ApJ...827...27Z} Zhang, Q.~M., Li, D., Ning, Z.~J., et al.\ 2016a, \apj, 827, 27

\bibitem[Zhang et al.(2016)]{2016ApJ...832...65Z} Zhang, Q.~M., Li, D., \& Ning, Z.~J.\ 2016b, \apj, 832, 65

\bibitem[Zheng et al.(2016)]{2016ApJ...823..136Z} Zheng, R., Chen, Y., \& Wang, B.\ 2016, \apj, 823, 136

\end{thebibliography}
\end{document}